\documentclass[aps,pre,reprint,superscriptaddress,showpacs,amsmath
]{revtex4-1}
 
\renewcommand{\vec}[1]{\boldsymbol{#1}}

\newcommand{\Eq}[1]{Eq.~(\ref{#1})}
\newcommand{\eq}[1]{Eq.~(\ref{#1})}
\newcommand{\Order}[1]{O(#1)}
\newcommand{\uvec}[1]{\boldsymbol{\hat{\textbf{#1}}}}
\newcommand{\Opt}[1]{\mathbf{#1}}

\begin{document}


\title{Propagating wave in the flock of self-propelled particles}


\author{Waipot Ngamsaad}
\email{waipot.ng@up.ac.th}
\affiliation{Division of Physics, School of Science, University of Phayao, Phayao 56000, Thailand}

\author{Suthep Suantai}
\affiliation{Department of Mathematics, Faculty of Science, Chiang Mai University, Chiang Mai 50200, Thailand}


\date{\today}

\begin{abstract}
We investigate the linearized hydrodynamic equations of interacting self-propelled particles in two dimensional space. It is found that the small perturbations of density and polarization fields satisfy the hyperbolic partial differential equations---that admit analytical propagating wave solutions. These solutions uncover the questionable traveling band formation in the flocking state of self-propelled particles. Below the critical noise strength, an unstable disordered state (random motion) undergoes a transient vortex and evolves to an ordered state (flocking motion) as unidirectional traveling waves. There appear two possible longitudinal wave patterns depending on the noise strength, including single band in stable state and multiplebands in unstable state. A comparison of theoretical and experimental studies is presented.
\end{abstract}


\maketitle


\section{Introduction}
The onset of collective motion can be found in various systems of the self-propelled objects, ranging from macromolecules, microorganism, animal, human, and swarming robots (see Ref. \cite{Vicsek2012} and references therein). The physics aspect of this phenomenon has been a current active research topic.

The minimal paradigm that can be used to describe this dynamics successfully is acknowledged to the Vicsek model \cite{Vicsek1995}. In this model, the point-like self-propelled particles (SPP) move at constant velocity in the direction of their orientation unit vector \cite{Vicsek1995}. The particles
interact locally by trying to align their directions of motion in presence of noise. If the noise strength is greater than the critical value, the SPP move in random direction. If the noise strength is below the critical value, the particles transit from random motion (disordered state) to \emph{flocking} (ordered state), where they form the \emph{coherence clusters} so that the individual members tend to move together in the same direction. The Vicsek model can be viewed as the flying XY spin model where the phenomenological hydrodynamic equations have been proposed for description at a continuum level by Toner and Tu \cite{Toner1995, Toner1998}. Recently, the hydrodynamic equations of SPP were derived from specified individual-based dynamics using several coarse-gaining frameworks, such as the Smoluchowski equation \cite{Baskaran2008a, Baskaran2008b, Bertin2015} and the Boltzmann equation \cite{Bertin2006, Bertin2009, Peshkov2012, Bertin2015}.

Based on hydrodynamic theory, the long wavelength mode fluctuations of density and velocity fields propagating as sound waves in SPP had been predicted by Tu and Toner \cite{Tu1998}. Later, the moving bands of the ordered state in the disordered state background were obviously found in the large-scale simulations of SPP by different researchers \cite{Gregoire2004, Chate2008, Bertin2009, Ginelli2010}. Two distinct robust pattern forms of the traveling waves in SPP including solitary moving band \cite{Bertin2009, Gopinath2012, Ihle2013} and moving multistripes \cite{Mishra2010} have been explored. Apart from wave patterns, fluctuating flocking states \cite{Mishra2010} and stationary radially symmetric asters \cite{Gopinath2012} in SPP have also been presented. In the experiments, the existence of traveling bands in active matter has been demonstrated for the system of actin filaments \cite{Schaller2010polar} and colloidal rollers \cite{Bricard2013emergence}. Recently, the linear sound wave in a system of colloidal rollers has been explored experimentally \cite{Geyer2018sounds}.

Analytical work has been carried out in order to gain deep insight into wave propagation dynamics in SPP. The standard method is the linear stability analysis of the hydrodynamic equations \cite{Bertin2009, Mishra2010, Gopinath2012}. Several authors agree that the traveling waves in SPP owe their emergence to instability of the homogeneous states \cite{Bertin2009, Mishra2010, Gopinath2012}. However, the linear stability analysis provides only the \emph{dispersion relation} \cite{Bertin2009} that is inadequate to characterize the spatiotemporal wave patterns. In a more rigorous study, the propagative \emph{Ansatz}, in which the wave profile travels along a direction with constant speed, has been postulated to be a solution for hydrodynamic equations of SPP in one-dimensional space (1D) \cite{Bertin2009, Mishra2010, Caussin2014, Solon2015b}. This approach reduces the hydrodynamic equations from nonlinear partial differential equations (PDEs) to nonlinear ordinary differential equations (ODEs). The nonlinear ODEs can be recast further into equivalent Newton's equation of motion for single particle moving in a potential field in the presence of friction by using a dynamical framework \cite{Caussin2014, Solon2015b}. This approach seems likely to classify the three different types of propagating patterns in SPP successfully, including solitary wave, multistripes wave and polar-liquid droplet. Nonetheless, solving the exact wave profiles explicitly by using this framework is challenging due to its nonlinearity. 

As classified in the textbook of Whitham \cite{Whitham2011linear}, there are two classes of wave solutions for the linear or nonlinear PDEs which consist of \emph{hyperbolic wave} and \emph{dispersive wave} solutions. The difference is that the hyperbolic wave propagates in two opposite directions along an arbitrary axis, a case in which the speeds need not be equal \cite{Webster1950, MyintU1987, Guenther1996}. Obviously, the previous analyses rely on the dispersive wave solution that propagates in one direction \cite{Bertin2009, Mishra2010, Gopinath2012, Caussin2014, Solon2015b}. In contrast to this, we deduced that the linearized hydrodynamic equations of SPP can formulate the hyperbolic-type PDEs. Therefore, the previous propagating wave assumption, which belongs to the dispersive wave solution \cite{Bertin2009, Mishra2010, Gopinath2012, Caussin2014, Solon2015b}, is still an incomplete wave feature of SPP.

In this work, we investigate the linearized hydrodynamic equations of SPP, which can be combined into the linear wave PDEs \cite{Webster1950, MyintU1987, Guenther1996}. Instead of performing the conventional mode analysis \cite{Bertin2009, Mishra2010, Gopinath2012}, we solve for the exact space and time dependent solutions of these equations by using the Riemann method \cite{Webster1950, MyintU1987, Guenther1996}. These linear analytical solutions are capable of capturing the dynamics of SPP in the vicinity of early and final states of the system. Especially, they can be used to classify the wave pattern formation in the flocking state of SPP clearly. 

\section{Hydrodynamic equations}
In this study, we consider a particular variant Vicsek model that has been studied by Farrell \emph{et al.} \cite{Farrell2012}. The advantage of this variant model is that it can explicitly map the microscopic physical parameters to the hydrodynamic equations through a coarse-grained process. Adapted from Ref. \cite{Farrell2012}, the hydrodynamic equations that describe evolution of particle number density field $\rho(\vec{r},t)$ and polarization field $\vec W(\vec r,t)$ in two-dimensional space (2D), are given by 
\begin{eqnarray}
\label{eq:den}
\rho_t &=& -v_0\nabla\cdot\vec{W}, \\
\label{eq:polar_const}
\vec{W}_t &=&  - \frac{v_0}{2}\nabla\rho + \left(\frac{\gamma}{2}\rho - \varepsilon -\frac{\gamma^2}{8\epsilon} |\vec{W}|^2\right)\vec{W} \nonumber\\
&& - \frac{3\gamma}{16\varepsilon}v_0\left(\vec{W}\cdot\nabla\right)\vec{W} - \frac{5\gamma}{16\varepsilon}v_0\vec{W}\left(\nabla\cdot\vec{W}\right) \nonumber\\
&& + \frac{5\gamma}{32\varepsilon}v_0\nabla\left(|\vec{W}|^2\right) + \frac{v_0^2}{16\varepsilon}\nabla^2\vec{W},
\end{eqnarray}
where $v_0$ is the moving speed of the particle, $\epsilon$ describes noise strength and $\gamma$ describes the strength of alignment. The polarization field is associated with the particle velocity field $\vec V(\vec r,t)$ in such a way that $v_0 \vec W(\vec r,t) = \rho(\vec{r},t) \vec V(\vec r,t)$. 
These equations are coarse-gained dynamics of an $N$ point-like SPP system in which particles move at constant speed $v_0$ in the direction of their orientation unit vector and interact with the vicinity or neighborhood via noisy alignment rule \cite{Vicsek1995}. The position $\vec{r}_i(t)$ and the orientation angle $\theta_i(t)$ of the $i$th particle at time $t$ evolve with the following equations of motion: 
$\dot{\vec{r}}_i = v_0\uvec{p}_i$ and $\dot{\theta}_i = \sum_{j\neq i} F(\theta_i-\theta_j,\vec{r}_i-\vec{r}_j)+\sqrt{2\varepsilon}\eta_i(t)$,
where the unit vector $\uvec{p}_i(\theta_i) = \cos\theta_i \uvec{x} + \sin\theta_i \uvec{y}$ and $\eta_i(t)$ is a white noise with zero mean and unit variance. The local pairwise alignment interaction is given by $F(\theta,\vec{r}) = \gamma\sin(\theta)/(\pi l^2)$, if $|\vec{r}|\leq l$ (otherwise $F=0$), where $l$ is the interaction range \cite{Farrell2012}. Although differences in physical parameters, \eq{eq:den} and \eq{eq:polar_const} have a form identical to the phenomenological model proposed by Toner and Tu \cite{Toner1995, Toner1998} and the coarse-grained equations obtained by using the Boltzmann theory \cite{Bertin2009}. 

The homogeneous states of \eq{eq:den} and \eq{eq:polar_const} admit arbitrary constant density $\rho_0$ with two possible values of polarization $W_0$, given by 
\begin{equation}
|\vec{W}_0| = W_0 = 
 \begin{cases} 
  0, & \varepsilon \geq \varepsilon_0\\ 
  \sqrt{8\varepsilon\left(\varepsilon_0-\varepsilon\right)}/\gamma, & \varepsilon < \varepsilon_0
 \end{cases} 
\end{equation} 
where $\varepsilon_0 = \gamma\rho_0/2$, which is defined as the critical noise strength value. Above the critical point ($\varepsilon > \varepsilon_0$), the system is in disordered state with zero polarization where the SPP move in random direction. Below the critical point ($\varepsilon < \varepsilon_0$), the system transitions into an ordered state where the SPP tend to move together in the same direction with nonzero polarization, called flocking. 


\section{Linearized wave equations}
Now we study the dynamics of SPP in the vicinity of homogeneous states. We suppose that the homogeneous polarization aligns in x-direction. Thus,  we define the solutions as follows: $\rho(\vec{r},t) = \rho_0 + n(\vec{r},t)$ and $\vec{W}(\vec{r},t) = W_0\vec{\hat x} + \vec{u}(\vec{r},t)$, where $n(\vec{r},t)$ and $\vec{u}(\vec{r},t)$ are small perturbations in density and polarization fields, respectively, called perturbations for convenience. Substituting these solutions into \eq{eq:den} and \eq{eq:polar_const} by first retaining the first-order terms and then obtaining the linearized hydrodynamic equations of SPP,
\begin{eqnarray}
\label{eq:den_lin}
n_t &=& -v_0\nabla\cdot\vec{u}, \\
\label{eq:polar_lin}
\vec{u}_t &=&  -\frac{v_0}{2}\nabla n + \alpha_0\vec{u} + \frac{\gamma}{2}W_0\vec{h}  + \frac{v_0^2}{16\varepsilon}\nabla^2\vec{u},
\end{eqnarray}
where $\alpha_0 = (\varepsilon_0-\varepsilon-\frac{\gamma^2}{8\varepsilon}W_0^2)$ and $\vec{h} = n\vec{\hat x} -\frac{\gamma}{2\varepsilon}W_0\left(\vec{\hat x}\cdot\vec{u}\right)\vec{\hat x} - \frac{v_0}{8\varepsilon} [ 3\left(\vec{\hat x}\cdot\nabla\right)\vec{u} - 5\vec{\hat x}\left(\nabla\cdot\vec{u}\right) + 5\nabla\left(\vec{\hat x}\cdot\vec{u}\right) ]$. The vector field $\vec{h}$ tends to drive the polarization field to the mean direction and affects only the flocking state ($W_0 \neq 0$). Operating \eq{eq:den_lin} with $\partial_t$ and using \eq{eq:polar_lin} (similarly, operating \eq{eq:polar_lin} with $\partial_t$ and using \eq{eq:den_lin}), we arrive at the following equations.
\begin{eqnarray}
\label{eq:den_lin_wave}
n_{tt} - \alpha_0 n_t &=& c^2\nabla^2 n - \frac{\gamma}{2}W_0 v_0 \nabla\cdot\vec{h} + \Order{\kappa\nabla^2 n_t}, \\
\label{eq:polar_lin_wave}
\vec{u}_{tt} - \alpha_0 \vec{u}_t  &=& c^2\nabla^2 \vec{u} + c^2\nabla\times\left(\nabla\times\vec{u}\right) + \frac{\gamma}{2}W_0\vec{h}_t \nonumber\\
&& + \Order{\kappa\nabla^2 \vec{u}_t},  
\end{eqnarray}
where $c = \frac{v_0}{\sqrt 2}$ and $\kappa = \frac{v_0^2}{16\varepsilon}$. Noting that the third-order derivative terms in \eq{eq:den_lin_wave} and \eq{eq:polar_lin_wave} can be ignored as we are interested only in the evolution of large flocking cluster or long wavelength ($\lambda$) mode, $\lambda \gg \frac{\pi v_0}{2\sqrt{|\varepsilon\alpha_0|}}$. Obviously, \eq{eq:den_lin_wave} and \eq{eq:polar_lin_wave} belong to the wave equations or hyperbolic PDEs \cite{Webster1950, MyintU1987, Guenther1996}.  

\subsection{Perturbation of disordered state}
From \eq{eq:den_lin_wave} and \eq{eq:polar_lin_wave}, the perturbations of the disordered state ($W_0=0$) satisfy the telegraph equations \cite{Webster1950, MyintU1987, Guenther1996} 
\begin{eqnarray}
\label{eq:den_telegraph}
n_{tt} - \alpha_0 n_t &=& c^2\nabla^2 n, \\
\label{eq:polar_telegraph}
\vec{u}_{tt} - \alpha_0 \vec{u}_t  &=& c^2\nabla^2 \vec{u} + c^2\nabla\times\left(\nabla\times\vec{u}\right).
\end{eqnarray}
Specially, \Eq{eq:polar_telegraph} reveals a vortex, defined by $\vec{\omega} = \nabla \times \vec{u}$. The governing equation for the perturbed vorticity was obtained by taking the curl operator ($\nabla \times$) to \eq{eq:den_lin} with $W_0=0$,
\begin{equation}\label{eq:vortex_disorder}
\vec{\omega}_t =  \left(\varepsilon_0-\varepsilon\right)\vec{\omega} + \kappa\nabla^2 \vec{\omega}.
\end{equation} 
By using the following solutions $n = e^{\frac{1}{2}\left(\varepsilon_0-\varepsilon\right) t} \tilde{n}$ and $\omega = e^{\left(\varepsilon_0-\varepsilon\right) t} \tilde{\vec\omega}$, we found that $\tilde{n}$ satisfies a simple linear wave equation $\tilde{n}_{tt} = c^2\nabla^2 \tilde{n} + \frac{1}{4}\left(\varepsilon_0-\varepsilon\right)^2 \tilde{n}$, which looks similar to the Klein-Gordon equation, while $\tilde{\vec\omega}$ satisfies the diffusion equation $\tilde{\vec\omega}_t = \kappa\nabla^2 \tilde{\vec\omega}$ \cite{MyintU1987}. So that, $c$ is interpreted as the speed of sound in disordered phase that has the magnitude of about 0.707 of the individual particle velocity $v_0$ \cite{Ihle2013}. And, $\kappa$ is diffusion constant. It implies that the disordered state is unstable below the critical point ($\varepsilon < \varepsilon_0$) and it evolves to the ordered state to form the flocking state. The transient vortex can be observed in the early stages of simulations of the Vicsek-type model \cite{Czirok1997spontaneously}. However, the vortex in the experimental system such as colloidal rollers seems robust due to the effect of additional repulsive interaction and confinement \cite{Bricard2015emergent}, which have been excluded in our investigation.

\subsection{Perturbation of ordered state}
As shown in \eq{eq:den_lin_wave} and \eq{eq:polar_lin_wave}, the perturbations around the flocking or the ordered state ($W_0 \neq 0$) tend to be biased to the mean direction by vector field $\vec{h}$. It is observed, at least in simulations, that the moving bands are unidirectional waves \cite{Gregoire2004, Chate2008, Bertin2009, Ginelli2010, Caussin2014, Solon2015b}. We write such a symmetry-broken field as $\vec{u}(\vec{r},t) = w(\vec{r},t) \uvec{x} + v(\vec{r},t) \uvec{y}$, where $w$ and $v$ are x- and y-component of the small perturbed polarization field, respectively. Now, we consider the longitudinal mode, where the wave profiles propagate in the same direction as that of mean polarization ($n_y = w_y = v_y = 0$). From \eq{eq:den_lin_wave} and \eq{eq:polar_lin_wave}, the wave equations in this case are provided by
\begin{eqnarray}
\label{eq:telegraph_den_order}
n_{tt} + \alpha n_t &=&  c^2 n_{xx} - 2\nu n_{tx} - \beta n_x + \Order{\kappa n_{txx}}, \\
\label{eq:telegraph_polar_order}
w_{tt} + \alpha w_t &=&  c^2 w_{xx} - 2\nu w_{tx}  - \beta w_x + \Order{\kappa w_{txx}}, 
\end{eqnarray}
where $\alpha = 2\left(\varepsilon_0-\varepsilon\right)$, $\beta = \frac{\gamma }{2}W_0v_0$, and $\nu = \frac{3\gamma}{32\varepsilon}W_0v_0$. Noting that $v$ is decoupled and tends to eventually decay to small value by a bias-diffusion process. 

Next, we consider the transverse mode where the wave profiles propagate perpendicular to the direction of mean polarization ($n_x = w_x = v_x = 0$). From \eq{eq:den_lin_wave} and \eq{eq:polar_lin_wave}, the wave equations for the ordered state in this case are given below.
\begin{eqnarray}
\label{eq:den_trans_wave}
n_{tt} &=& c^2 n_{yy} + \Order{\kappa n_{txx}}, \\
\label{eq:polar_trans_wave}
v_{tt} &=& c^2 v_{yy} + \Order{\kappa v_{txx}}.
\end{eqnarray}
$w$ is assumed to relax to zero in the ordered state. By neglecting the third-order derivative term, \Eq{eq:den_trans_wave} and \eq{eq:polar_trans_wave} are the plane wave equations with the well-known d'Alembert solution---where the initial condition splits into two waves that propagate in opposite directions along the y-axis with the speed of sound $c$ \cite{Webster1950, MyintU1987, Guenther1996}. Since the perturbations do not change shape from the initial conditions for this sort of wave, we ignore this mode in the present study. In addition, the speed of sound in the experimental system of colloidal rollers is direction-dependent \cite{Geyer2018sounds}. It results from the hydrodynamic and electrostatic interactions in the system of colloidal rollers \cite{Bricard2013emergence,Geyer2018sounds} that have been excluded in our model. 

We are now looking for the analytical space and time dependent solution of longitudinal waves. By dropping the third-order derivative term, we rewrite \eq{eq:telegraph_den_order} or \eq{eq:telegraph_polar_order} 
\begin{equation}
\label{eq:telegraph_polar_solve1}
\phi_{tt} + 2\nu \phi_{tx} - c^2 \phi_{xx} + \alpha \phi_t + \beta \phi_x = 0,
\end{equation}
where $\phi$ can refer to either $n$ or $w$, since all the equations are in identical form. The initial conditions for \eq{eq:telegraph_polar_solve1} are given by $\phi(x,0) = f(x)$, $D_t\phi(x,0) = \phi_t(x,0) + \nu\phi_x(x,0) \equiv g(x)$. \Eq{eq:telegraph_polar_solve1} is a second-order PDE whose characteristic equation is given by $\left(\frac{dx}{dt}\right)^2 - 2\nu\left(\frac{dx}{dt}\right) - c^2 = 0$ or $\frac{dx}{dt} = \nu\pm\sqrt{\nu^2+c^2}$ \cite{Webster1950, MyintU1987, Guenther1996}. From the characteristic equation, obviously, \Eq{eq:telegraph_polar_solve1} is a hyperbolic-type PDE and it can be reduced to a \emph{canonical form} by introducing the curvilinear coordinates given below.
\begin{equation}\label{eq:char_var}
\eta = x + c^{-}t,\qquad \xi = x - c^{+}t,
\end{equation}
where $c^\pm = \sqrt{\nu^2+c^2} \pm \nu$, so that $\nu$ is exactly the collective speed of SPP induced by the alignment interaction. Applying the transformations in \eq{eq:char_var}, we rewrite \eq{eq:telegraph_polar_solve1} in $\eta\xi$-plane 
\begin{equation}\label{eq:reduce_wave}
\phi_{\eta\xi} + k^{-} \phi_\eta +k^{+} \phi_\xi = 0,
\end{equation}
where $k^{-} = -\frac{\alpha c^{-}+\beta}{4\Lambda^2}$, $k^{+} = \frac{\alpha c^{+}-\beta}{4\Lambda^2}$, and $\Lambda = \sqrt{\nu^2 + c^2} = \frac{1}{2}(c^- + c^+)$. Now the solutions of \eq{eq:reduce_wave} depend on the two wave variables, $\phi(x,t) = \phi(\eta,\xi)$. According to $\nu > 0$ in the ordered state, the wave speeds $c^\pm$ are always positive so that $\eta$ and $\xi$ are left- and right-propagating wave variables, respectively. In the presence of collective motion, the wave speeds in the flocking state are larger than the speed of sound in the disordered phase. This supports the \emph{supersonic wave} structure as pointed out by Ihle \cite{Ihle2013}. 

\section{Riemann method}
Finding the solution of \eq{eq:reduce_wave} subjected to the initial data is called a \emph{Cauchy problem}, which can be solved by using the \emph{Riemann method} \cite{Webster1950, MyintU1987, Guenther1996}. This approach can solve the general form of linear hyperbolic PDE in 1D, but case study in the presence of $\nu$ term is scarce \cite{Webster1950, MyintU1987, Guenther1996}. Therefore, we provide the procedure for solving \eq{eq:reduce_wave} as follows. 

For convenience in further calculation, we rewrite \eq{eq:reduce_wave}
\begin{equation}\label{eq:Cauchy1}
\Opt{L}[\phi] = \phi_{\eta\xi} + k^{-} \phi_\eta +k^{+} \phi_\xi = 0,
\end{equation}
where $\Opt{L}$ is linear operator. From \eq{eq:char_var}, we have that $t = \frac{\eta-\xi}{c^{-}+c^{+}}$ and $x = \frac{c^{+}\eta+c^{-}\xi}{c^{-}+c^{+}}$. When $t=0$, we have $\eta=\xi=x$, which is the straight line in the $\eta\xi$-plane. Therefore, the initial conditions in $\eta\xi$-plane (\emph{Cauchy data}) are transformed to
\begin{eqnarray}
\label{eq:init_ux_tran}
\phi\vert_{\eta=\xi} &=& f(\xi), \\
\label{eq:init_dux_tran}
\Opt{M}[\phi]\vert_{\eta=\xi} &=& \frac{1}{\Lambda}\left(\phi_t(\xi) + \nu \phi_x(\xi)\right) \equiv \frac{1}{\Lambda} g(\xi),
\end{eqnarray}
where we define operator $\Opt{M}[*] = \partial_\eta * - \partial_\xi *$. 

The starting point of the Riemann method is to find a smooth function $R(\eta,\xi)$, called the \emph{Riemann function}, that satisfies the adjoin equation
\begin{equation}\label{eq:riemann2}
\Opt{L}^*[R] = R_{\eta\xi} -k^{-} R_\eta - k^{+} R_\xi = 0,
\end{equation}
where $\Opt{L}^\ast$ is \emph{adjoin operator} \cite{Webster1950, MyintU1987, Guenther1996}. This approach can reduce the second-order PDE to the first-order integral equation. From \eq{eq:Cauchy1} and \eq{eq:riemann2} it evaluates that
\begin{equation}\label{eq:riemann3}
R\Opt{L}[\phi] - \phi\Opt{L}^\ast[R] = P_\eta + Q_\xi = 0,
\end{equation}
where $P = \frac{1}{2}\left(R \phi_\xi - \phi R_\xi \right) + k^{-}R\phi$ and $Q = \frac{1}{2}\left(R \phi_\eta - \phi R_\eta\right) + k^{+}R\phi$. By using Green's theorem \cite{Arfken1985, Webster1950, MyintU1987, Guenther1996} in \eq{eq:riemann3}, we have $ \iint_D\left(P_\eta + Q_\xi\right)d\eta d\xi = \oint_C\left(Pd\xi-Qd\eta\right) = 0$, where $D$ is the region bounded by the positively oriented closed curve $C$. We integrate along the three edges of a triangle in $\eta\xi$-plane whose vertices with positive orientation are given by $C_0 = (\eta_0,\xi_0)$, $C_1 = (\eta_0,\eta_0)$ and $C_2 = (\xi_0,\xi_0)$. In this way, we choose a path that $d\eta=0$ along $C_0\textnormal{-}C_1$ line, $d\xi=0$ along $C_2\textnormal{-}C_0$ line and $d\eta=d\xi$ along $C_1\textnormal{-}C_2$ line, containing the initial data in \eq{eq:init_ux_tran} and \eq{eq:init_dux_tran}. So that, the Riemann method turns our problem to a line integral equation 
\begin{equation}\label{eq:curve_int}
\int_{C_0}^{C_1}Pd\xi + \int_{C_1}^{C_2}\left(P-Q\right)\vert_{\eta=\xi}d\xi -\int_{C_2}^{C_0}Qd\eta = 0.
\end{equation}
To calculate the integral terms in \eq{eq:curve_int}, the Riemann function must  satisfy following conditions: $R_\xi - k^{-}R = 0$ when $\xi = \xi_0$, $R_\eta - k^{+}R = 0$ when $\eta = \eta_0$ and $R = 1$ when $\eta=\eta_0$ and $\xi = \xi_0$. After evaluating \eq{eq:curve_int} with the properties of Riemann function and initial data \eq{eq:init_ux_tran} and \eq{eq:init_dux_tran}, we have 
\begin{widetext}
\begin{equation}\label{eq:sol}
\phi(\eta_0,\xi_0) = \frac{1}{2}\left[R(\eta_0,\eta_0) f(\eta_0) + R(\xi_0,\xi_0) f(\xi_0) \right]  - \frac{1}{2}\int_{\eta_0}^{\xi_0} \left\lbrace R(\xi,\xi) \Opt{M}[\phi]\vert_{\eta=\xi} - \Opt{M}[R]\vert_{\eta=\xi} f(\xi) + 2\Gamma R(\xi,\xi) f(\xi) \right\rbrace d\xi,
\end{equation}
\end{widetext}
where $\Gamma = k^+ - k^-$. \eq{eq:sol} is analytical solution of our main problem in $\eta\xi$-plan. The remain ingredient is the exact form of the Riemann function.


\subsection{Riemann function}
To find the Riemann function, we define $R(\eta,\xi) = \exp{\left[k^{+}\left(\eta-\eta_0\right) + k^{-}\left(\xi-\xi_0\right)\right]}\Psi(\eta,\xi)$. Substituting it to \eq{eq:riemann2}, we obtain $\Psi_{\eta\xi} - \frac{k^2}{4} \Psi = 0$, where $k^2 = 4k^{-}k^{+}$. We assume that this equation has a particular solution in the form $\Psi(q)$ where $q = \left(\eta-\eta_0\right)\left(\xi-\xi_0\right)$ \cite{Webster1950, MyintU1987, Guenther1996}. Using this assumption, we have $q \ddot{\Psi}(q) + \dot{\Psi}(q) - \frac{k^2}{4} \Psi(q) = 0$. This equation can be transformed further with another new variable $\theta = k\sqrt{q}$ and then we have $\ddot{\Psi}(\theta) + \frac{1}{\theta}\dot{\Psi}(\theta) - \Psi(\theta) = 0$. Finally, it found that $\Psi(\theta)$ exactly satisfies the zeroth-order modified Bessel equation whose solution has been known. After gathering all terms, the Riemann function is provided by 
\begin{equation}\label{eq:Riemann_func}
R(\eta,\xi) =  A(\eta,\xi) I_0\left(k\sqrt{\left(\eta-\eta_0\right)\left(\xi-\xi_0\right)}\right),
\end{equation}
where $A(\eta,\xi) = \exp{\left[k^{+}\left(\eta-\eta_0\right) + k^{-}\left(\xi-\xi_0\right)\right]}$ and $I_0$ is the zeroth-order modified Bessel function \cite{Arfken1985}. It can calculate that $R_\eta = k^{+}R + \frac{k^2}{2}\left(\xi-\xi_0\right) A(\eta,\xi) \frac{I_1(\theta)}{\theta}$ and $R_\xi = k^{-}R  + \frac{k^2}{2}\left(\eta-\eta_0\right) A(\eta,\xi) \frac{I_1(\theta)}{\theta}$, where $I_1(\theta) = \dot{I}_0(\theta)$ which is first-order modified Bessel function \cite{Arfken1985}. And it shows that this Riemann function satisfies all required conditions.

As shown later, we transform the solution in \eq{eq:sol} back to $xt$-plane by using the Riemann function \eq{eq:Riemann_func} subjected to the initial data \eq{eq:init_ux_tran} and \eq{eq:init_dux_tran}. Since $\xi$ becomes dummy variable now, we let $\eta_0 = x + c^{-}t$ and $\xi_0 = x - c^{+}t$. 

\subsection{Stability analysis}
We now find stability of the obtained analytical solution by considering the exponential factor in \eq{eq:Riemann_func} along $\eta=\xi$ line that $A(\xi,\xi) = \exp{\left[k^{+}\left(\xi-\eta_0\right) + k^{-}\left(\xi-\xi_0\right)\right]}$. For $\varepsilon < \varepsilon_0$, it found that $k^- < 0$, due to $c^\pm$, $\alpha$ and $\beta$ are always positive, while $k^+$ can be either negative or positive. Solving the inequality, we found that $k^+ < 0$ if $\varepsilon > \frac{7}{11}\varepsilon_0$ and $k^+ > 0$ if $\varepsilon < \frac{7}{11}\varepsilon_0$. For $\xi_0 \leq \xi \leq \eta_0$ or equivalent to $-c^- t \leq x-\xi \leq c^+ t$, therefore $A(\xi,\xi)$ always decays when $0 < \varepsilon < \frac{7}{11}\varepsilon_0$  (stable regime). In contrast, $A(\xi,\xi)$ can grow when $\frac{7}{11}\varepsilon_0 < \varepsilon < \varepsilon_0$ (unstable regime). The growth rate is highest at $\xi=\xi_0$ and trends to decrease as $\xi<\xi_0$. Consequently, $k^2 > 0$ if $\frac{7}{11}\varepsilon_0 < \varepsilon < \varepsilon_0$ (unstable regime) and $k^2 < 0$ if $0 < \varepsilon < \frac{7}{11}\varepsilon_0$ (stable regime). Using the relation $I_m(s) = i^{-m}J_m(is)$ where $J_m(s)$ is the Bessel function of order $m$, the Riemann function for $k^2 < 0$ changes to $R(\eta,\xi) =  A(\eta,\xi) J_0\left(k\sqrt{\left(\eta-\eta_0\right)\left(\xi-\xi_0\right)}\right)$.

Gartering all terms, the analytical wave solution of \eq{eq:reduce_wave} in space and time variables is provided by 
\begin{widetext}
\begin{equation}\label{eq:fin_sol_real}
\phi(x,t) = \frac{1}{2}\left[e^{a^- t}f(x + c^{-}t) + e^{a^+ t}f(x - c^{+}t)\right] + \frac{1}{2} e^{-\left(\mu x - \sigma t\right)} \int_{x - c^{+}t}^{x + c^{-}t} e^{\mu \xi} \left[F(x-\xi,t)f(\xi) + G(x-\xi,t)g(\xi)\right] d\xi,
\end{equation}
\end{widetext}
where $a^- = 2\Lambda k^{-}$, $a^+ = -2\Lambda k^{+}$, $\mu = k^+ + k^-$, and $\sigma = k^{-}c^{+} - k^{+}c^{-}$. For $-c^- t < x < c^+ t$ and $4k^-k^+ = k^2 > 0$, the propagators $F$ and $G$ are given by 
\begin{eqnarray}
\label{eq:propagator_F_unstable}
F(x,t) &=& - \Gamma J_0(k s(x,t))  + \Lambda k t \frac{J_1(k s(x,t))}{s(x,t)}, \\
\label{eq:propagator_G_unstable}
G(x,t) &=& \frac{1}{\Lambda} J_0(k s(x,t)),
\end{eqnarray}
where $s(x,t) = \sqrt{c^2t^2 + 2\nu xt - x^2}$ and $\Gamma = k^+ - k^-$. The wave profile in \eq{eq:fin_sol_real} is nonzero for the position from $x - c^- t$ to $x + c^+ t$, called \emph{domain of dependence} \cite{MyintU1987, Guenther1996}, that supports the finite band formation and discontinuous front as found in simulations \cite{Chate2008, Bertin2009, Ihle2013}. 

\section{Discussion}
From \eq{eq:fin_sol_real}, the analytical solution indicates that initial profiles of the small perturbed density and polarization fields lose their configuration and propagate in both positive and negative directions of x-axis with unequal speed. Due to $c^+ > c^-$, the propagation in the positive direction is faster than in the negative. Below the critical point, $\varepsilon < \varepsilon_0$, we found that $k^+ > 0$ when $\varepsilon < \frac{7}{11}\varepsilon_0$ and $k^+ < 0$ when $\varepsilon > \frac{7}{11}\varepsilon_0$ while $k^-$ is always negative. As $t \gg 0$, according to $\Lambda > 0$, the left-propagating initial profile decays to zero whereas the right-propagating wave grows for $\varepsilon > \frac{7}{11}\varepsilon_0$ (unstable regime) and decays for $\varepsilon < \frac{7}{11}\varepsilon_0$ (stable regime). Therefore, the propagating waves in SPP trend move in the direction of mean polarization vector as found in simulations \cite{Gregoire2004, Chate2008, Bertin2009, Ginelli2010, Caussin2014, Solon2015b}.

To this point, there exists another transition noise strength at $\frac{7}{11}\varepsilon_0$ that separates the spatiotemporal pattern formation of the propagating wave in the flocking state of SPP into two regimes as mentioned by Chat\'e \emph{et al.} \cite{Chate2008}. Let us consider the unstable regime where $4k^-k^+ = k^2 > 0$. The asymptotic Bessel function is given by $J_m(ks) = \sqrt\frac{2}{\pi ks}\cos(ks - \frac{\pi}{4} - \frac{m\pi}{2})$ for $s \gg 1$. With this character, it shows that the perturbations of the unstable ordered state propagate as waves with spatial oscillatory pattern or multiplebands in 2D, that has been observed in simulations \cite{Chate2008, Mishra2010, Caussin2014, Solon2015b}. The wave profiles grow with the fastest rate in the position of the leading front and slower for the tandem position, at least in the early stage. Therefore, $k$ is equivalent to wavenumber which relates to the wavelength $\lambda_w$ as follows: $\lambda_w = \frac{2\pi}{k} = \frac{\pi}{\sqrt{k^{-}k^{+}}}$. Thus, we have
\begin{equation}\label{eq:bands_width}
\lambda_w = 2\pi \frac{\frac{9}{64}\left(\frac{1-\varepsilon^\prime}{\varepsilon^\prime}\right)+1}{\sqrt{\frac{11}{2}\left(\varepsilon^\prime-\frac{7}{11}\right)\left(1-\varepsilon^\prime\right)}}\frac{v_0}{\varepsilon_0} ,
\end{equation}
where $\varepsilon^\prime = \frac{\varepsilon}{\varepsilon_0}$. The wavelength in \eq{eq:bands_width} can be used to approximate the width of the stripes and it shows that the moving speed, $v_0$, of the particle, plays a role in regulating the bands width. In the opposite situation, for the stable regime where $4k^-k^+ = -k^2 < 0$, the propagators change to
\begin{eqnarray}
\label{eq:propagator_F_stable}
F(x,t) &=& - \Gamma I_0(k s(x,t))  - \Lambda k t \frac{I_1(k s(x,t))}{s(x,t)}, \\
\label{eq:propagator_G_stable}
G(x,t) &=& \frac{1}{\Lambda} I_0(k s(x,t)).
\end{eqnarray}
The asymptotic form of the modified Bessel functions, given by $I_m(ks) \sim \frac{1}{\sqrt{2\pi ks}}e^{ks}$ for $s \gg 1$, shows the nonoscillatory wave patterns or a single band in 2D. In the stable ordered state, the perturbations eventually decay to smaller values. Thus, our linear approximation should be valid over long time scale. In long time scale $t \rightarrow \infty$, we approximate $s \simeq ct + \frac{\nu}{c}x - \frac{1}{2}\frac{\Lambda^2}{c^3}\frac{x^2}{t}$. Thus, below the noise threshold, the perturbations converge to the homogeneous ordered state as biased Gaussian waves. However, the homogeneous ordered state is not observed in the simulation; instead, this state is replaced by the fluctuating flocking state \cite{Chate2008, Mishra2010}.

\section{Conclusion}
In summary, based on the linearized hydrodynamic theory of self-propelled particles, the small perturbed density and polarization fields are governed by the hyperbolic partial differential equations. As opposed to the previous analytical studies that rely on the dispersive wave solution, our analytical hyperbolic wave solutions reveal some different aspects of spatiotemporal pattern formations in self-propelled particles. Below the critical noise strength, the homogeneous disordered state is unstable that is growing into the ordered state and generates the vortex flow of perturbation polarization field. The perturbations in the homogeneous ordered state evolve as two possible unidirectional longitudinal propagating waves separated by a threshold noise strength. This includes single band in the stable state below the threshold value and multiplebands in the unstable state above the threshold value. We believe that these special solutions could provide the basic knowledge for studying the dynamics of more complex self-propelled particles, by including hydrodynamic and electrostatic interactions, in the future work.

\begin{acknowledgments}
This research was supported by the Research Grant for New Scholar (\emph{Grant no. MRG5980258}) funded by The Thailand Research Fund (TRF) and Office of the Higher Education Commission (OHEC). 
\end{acknowledgments}


\bibliography{DDRDE_ref}


\end{document}